\documentclass[12pt,preprint]{aastex}

\newcommand{\Tsun}{\mbox{$T_\odot$}}

\newcommand{\mearth}{\mbox{$M_\oplus$}}
\newcommand{\msun}{\mbox{$M_\odot$}}

\newcommand{\mic}{\mbox{$\mu$m}}
\newcommand{\app}{\mbox{$\sim$ }}
\newcommand{\rin}{\mbox{$R_{in}	$ }}
\newcommand{\delr}{\mbox{$\Delta R$ }}

\newcommand{\mjya}{\mbox{mJy/arcsec$^2$}}
\newcommand{\dg}{\mbox{$^\circ$}}

\begin{document}

\title {High-Resolution Imaging of the Dust Disk around 49 Ceti}

\author {Z. Wahhaj\altaffilmark{1,2}, 
D.W. Koerner\altaffilmark{2} and
A.I. Sargent\altaffilmark{3}}
\altaffiltext{1} {Institute for Astronomy, University of Hawaii, Honolulu, HI 96814}
\altaffiltext{2} {Northern Arizona University,  Building 19, Rm. 209, Flagstaff, AZ 86011-6010}
\altaffiltext{3} {Dept. of Astronomy, California Institute of Technology, Pasadena, CA 91125}


\begin{abstract}

Sub-arcsecond scale Keck images of 
the young A1V star, 49 Ceti, resolve emission at $\lambda = 12.5$ 
and $17.9$~\mic\ from a disk with long axis at {PA} $125\pm10$\dg and 
inclination $\phi= 60\pm15$\dg. At $17.9$ \mic, the emission
is brighter and more extended toward the NW than the SE.
Modeling of the mid-infrared
images combined with flux densities from the
literature indicate that the bulk of the mid-infrared emission comes 
from very small grains ($a$ \app0.1 \mic) confined between 
$30$ and $60$ AU from the star.  This population of dust 
grains contributes negligibly to the significant excess observed 
in the spectral energy distribution. Most of the non-photospheric 
energy is radiated at longer wavelengths by an outer disk of 
larger grains ($a$~\app15~\mic ), inner radius \app60 AU, 
and outer radius \app900 AU. Global properties of the 
49 Cet disk show more affinity with the $\beta$ Pic and
HR 4796A disks than with other debris disks. 
This may be because they are all very young ($t < 20$ Myr), 
adding strength to the argument that they are 
transitional objects between Herbig Ae and ``Vega-like''
A stars with more tenuous circumstellar disks.

\end{abstract}


\section {Introduction}

Vega-excess stars were identified in observations with the 
Infrared Astronomical Satellite (IRAS) as having infrared excess 
emission attributed to circumstellar dust (see reviews 
by Backman \& Paresce 1993 and Zuckerman 2001). More than a hundred 
Vega-excess candidates have been identified by comparison of IRAS 
sources with stellar counterparts in optical catalogs 
since the early discovery of prototypes such as Vega and 
$\beta$ Pic (Backman \& Paresce 1993, Mannings \& Barlow 1998, Silverstone 
2000). Estimated ages and circumstellar dust masses support the view 
that these objects are largely distinct from their pre-main sequence 
counterparts, the T Tauri and Herbig Ae/Be stars. The latter have typical 
ages less than 10$^7$ yrs and fractional infrared luminosities, 
$\tau = L_{ir}$/$L_{bol}$, of approximately 0.25 
(Adams, Lada \& Shu 1987), while Vega-excess stars may be as old as 10$^9$ 
yrs with values of $\tau$ typically in the range $10^{-4}$ to $10^{-7}$. 
Larger values of $\tau = 10^{-3}$ are found for a few Vega-excess stars
with much younger ages (t $\sim 10^7$ yrs) that often appear to be associated with 
young kinematic moving groups (Moor et al.\ 2006). Jura et al.\ (1998) noted 
three A-type stars in the Bright Star Catalogue with $\tau \sim 10^{-3}$ 
and interpreted them as representative of an early phase of debris 
disk evolution. $\beta$ Pic, HR 4796A, and 49 Cet have since been assigned 
stellar ages of 20 Myr or less (Stauffer et al.\ 1995, 
Zuckerman et al.\ 1995, Zuckerman et al.\ 2001) and may well be 
important as examples of the transition from viscous protostellar accretion 
disks to planetary systems. The existence of such a phase 
is supported by recent {\it Spitzer} studies of weakline T Tauri stars in 
nearby clouds where a small sample of objects with values of $\tau$ 
in the high debris disk range was recently discovered by Cieza et al.\ (2006).

High-resolution images can 
reveal features that signal the presence of gravitational
interactions between unseen planets and orbiting dust grains.
In recent years more than a dozen such disks have been resolved,
revealing tantalizing features like ring arcs, clumps, disk warps, \
and other non-axisymmetric signatures of potentially planetary importance 
(e.g. 
Holland et al.\ 1998, 
Koerner et al.\ 1998,
Schneider et al.\ 1999, 
Heap et al.\ 2000, 
Koerner et al.\ 2001, 
Wilner et al.\ 2002, 
Clampin et al.\ 2003, 
Wahhaj et al.\ 2003, 
Ardila et al.\ 2004, 
Liu et al.\ 2004,
Schneider, Silverstone \& Hines 2005, 
Wahhaj et al.\ 2005, 
Kalas et al.\ 2005 and 
Kalas, Graham \& Clampin 2006 
). 
Simple gaps and rings are often apparent, although
these may not be definitive signatures of unseen planets (Takeuchi \& 
Artymowicz 2001, Klahr \& Lin 2001). Some features, however, are as yet 
unaccounted for by any other means. Examples include stellar offset from 
the orbital center of an elliptical ring around Fomalhaut (Kalas et al.\ 
2005). The evidence for a planet-disk connection is now sufficiently compelling that the
Vega-excess phenomenon has been proposed as an indirect detection 
method for planetary systems (Zuckerman \& Song 2004).

Dynamically important disk features have been identified in high-resolution
mid-infrared images for 2 out of the three young A-type stars mentioned by 
Jura et al.\ (1998). Keck studies of HR 4796A (Koerner et al.\ 1998; 
Telesco et al\ 2000; Wahhaj et al.\ 2005) and $\beta$ Pic
(Wahhaj et al.\ 2003; Weinberger et al.\ 2003; see also Gemini mid-infrared 
observations by Telesco et al.\ 2005) reveal gaps and rings with 
azimuthal asymmetries, inclination offsets, and radial changes in grain size 
and distribution. Here we present Keck mid-infrared imaging of 49 Cet, the 
third nearby A-type star identified by Jura et al.\ (1998) as having a young 
debris disk in an early transitional phase. 49 Ceti is an A1V star with an age 
of 8$\pm$2 Myr (Zuckerman et al.\ 1995; Thi et al.\ 2000) and a 
$\it{Hipparcos}$ distance of 61$\pm$3 pc. As discussed by Jayawardhana 
et al.\ (2001), mid-infrared radiation of 49 Cet (0.38 Jy in the IRAS 
25 \mic\ band) is reduced in comparison to HR 4796A, posing an added 
challenge to imaging at mid-infrared wavelengths. Nevertheless, these 
authors report radiation from the circumstellar dust that is marginally 
resolved at 10.8 \mic\ with flux density $250\pm50$ mJy. Substantial
long-wavelength excess has also been detected at 
$\lambda$ = 60 \mic\ by Sadakane \& Nishida (1986) and at 
$\lambda$ = 1.2 mm ($F_{1.2mm}$ = 12.7 mJy) by Bockelee-Morvan 
et al.\ (1994). Here we present Keck II images of 49 Cet at $\lambda$ = 12.5 
and 17.9 \mic\ together with simultaneous modeling of both the images
and spectral energy distribution.

\section{Observations and Results}

49 Cet was imaged with the Mid-InfraRed Large-well Imager (MIRLIN; Ressler et al.\ 1994) 
at the f/40 bent-Cassegrain focus of the Keck II 
telescope on UT dates 7 October 1998  and 28-31 July 1999. 
At Keck, MIRLIN has a plate scale of 
0$''$.138 pixel$^{-1}$ and a 17$''$.5 field of view.  Filters centered
on $\lambda =$ 12.5 and 17.9 \mic\ with widths 1.16 and 2 \mic\ were
used for the observations presented here. Initial sky subtraction was 
carried out by differencing images chopped 7$''$ in an east-west direction
at a frequency of 4 Hz. The telescope was nodded the same distance
in a north-south direction after co-adding several hundred frames
at each of the chop positions. 
The total integration times were 20 and 66 minutes 
at 12.5 and 17.9 \mic\ respectively. 
Residual background emission in the form of a striping
pattern was removed by masking on-source emission and separately
subtracting the median emission within each
row and column of pixels. Disk emission
in each of the double-differenced chop and nod frames was shifted
and added to produce a final combined image. Observation and data
reduction of standards $\alpha$ Boo, $\alpha$ Tau, 
$\chi$ Cet and Vega proceeded
in the same way for photometric calibration and 
served as a representation of the Point Spread Function (PSF).
Images of 49~Cet and PSF stars from  separate nights of 
observation were co-added after 
weighting by the signal-to-noise ratio of their peak flux. 
Seeing effects on low-level emission in the PSF stars were 
mostly evident at 12.5~\mic\ and were 
reduced for both the source and calibration images
by smoothing with a circular hat function.

\begin{figure}
  \centerline{
    \includegraphics[height=8cm]{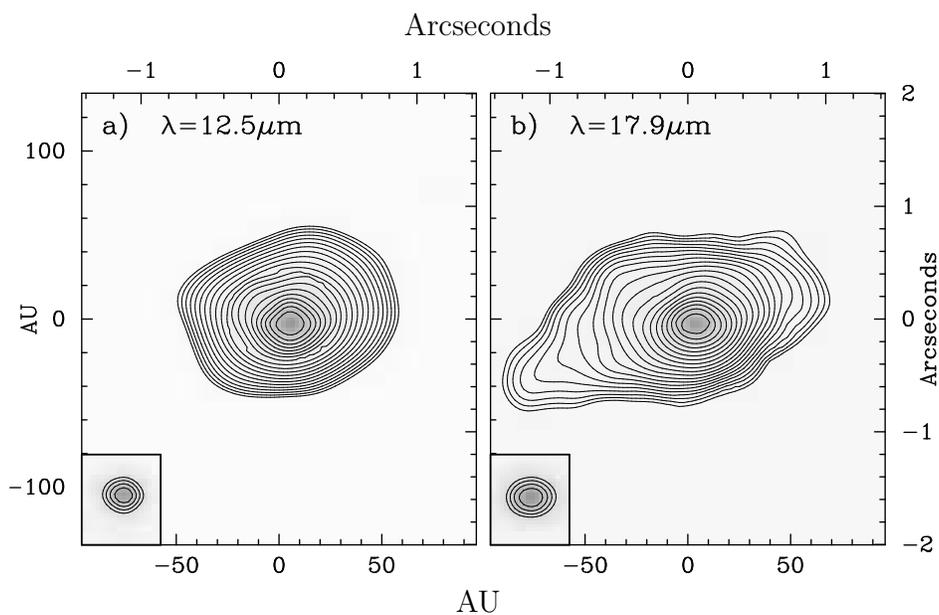}	
  }
  \caption{KECK/MIRLIN contour maps of 49 Cet at $\lambda =$ 12.5 
	and 17.9 \mic . North is ``up''.
	Contour images of the PSF star, $\alpha$ Boo are displayed
	in panels {\bf a} \& {\bf b} as insets with contour levels
  	of 10\%, starting at the 60\% level. The FWHM of the PSFs
	are 0$''$.41 and 0$''$.48 at 12.5 and 17.9 \mic\ respectively.
	{\bf (a)} Emission at 12.5 \mic . Lowest contour level is at the
	2$\sigma$ level (41 \mjya). Higher contour levels are at
	2$\sigma \times(10^{0.068n})$ for the $n^{th}$ contour.
	{\bf (b)} Emission at 17.9 \mic\ contoured as in {\bf a}. Lowest
	contour is at the 2$\sigma$ level (26 \mjya).} 
  \label{data}
\end{figure}

\begin{figure}
  \centerline{
    \includegraphics[height=8cm]{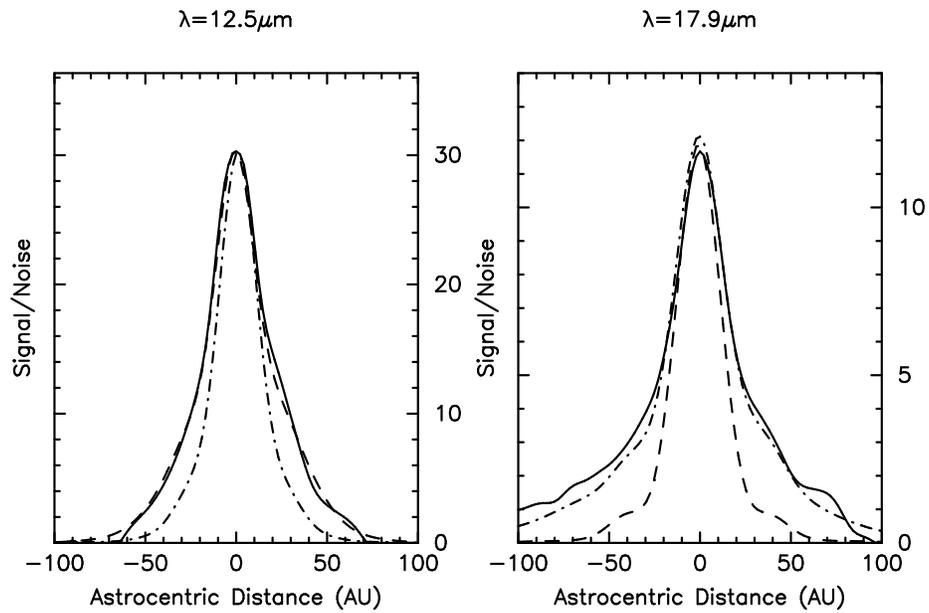}	
  }
  \caption{  Flux density strip cuts of width equal to FWHMs of PSFs taken 
        along the long axes of emission ( PA = 125\dg ) at 12.5 and 17.9\mic\ 
        seen in Fig.~1. 
	Solid lines represent the 49 Cet emission profiles and 
	dot-dashed lines the profile for PSF star $\alpha$ Boo.  
	Dashed lines are from best-fit models of emission 
	described in Section 3 of the text.}
  \label{scut}
\end{figure}

Final processed images of 49~Cet at angular resolutions
of 0$''$.41 and 0$''$.48 at $\lambda =$ 12.5 and 17.9 \mic, 
respectively are shown in Fig.~\ref{data}. The original 64 $\times$ 64 
pixel (8.8$''$ $\times$ 8.8$''$) images were finely gridded to 
256 $\times$ 256 pixels and smoothed by convolution
with a circular hat function of diameter equal to the 
Full Width Half Maximum (FWHM) of the PSF star $\alpha$ Boo. Images of 
$\alpha$ Boo, resampled and smoothed in the same manner, are shown
in the insets in Fig.~\ref{data}. 
Images used later 
for model-fitting purposes were smoothed in the same way,  
but not resampled beforehand. In Fig.~\ref{data} both 
images of 49~Cet are elongated along approximately 
the same position angle. To characterize this
effect more accurately, normalized strip cuts were obtained from the
images by windowing the emission in rectangular apertures 
aligned with the long axis of emission (PA = 125\dg ) and of 
width corresponding to the PSF FWHM.
The results, displayed in Fig.~\ref{scut}, demonstrate 
that 49 Cet's emission profile is clearly resolved at both
wavelengths.

Flux densities for 49~Cet were derived using aperture 
photometry of images of the calibrator
stars. Calibrator flux densities were adopted from 
the ISO Calibration web site 
(http://www.iso.vilspa.esa.es/users/expl\_lib/ISO/wwwcal/) as follows:
$\alpha$ Boo ($F_{12.5} = 478$ Jy; $F_{17.9} = 232$ Jy), 
$\alpha$ Tau ($F_{12.5} = 423$ Jy; $F_{17.9} = 205$ Jy), 
$\chi$ Cet ($F_{12.5} = 0.68$ Jy; $F_{17.9} = $0.33 Jy) 
and Vega ($F_{12.5} = 24.9$ Jy; $F_{17.9} = 12.2$ Jy). 
The resulting flux densities for 49~Cet
were 200$\pm$26 mJy and 186$\pm$25 mJy at 12.5 and 17.9 \mic.
Although our 12.5 \mic\ value and the appreciably higher 
IRAS 12 \mic\ measurement (330 mJy) are at first sight 
inconsistent, the discrepancy is explained readily by 
the fact that the IRAS band is much broader (6 to 18 \mic) and thus 
includes more of the bright photospheric emission at 
shorter wavelengths. Peak brightnesses, the flux in the 
brightest pixel divided by the pixel area, were 
231 \mjya\ and 583 \mjya\ at 
12.5 and 17.9 \mic\ respectively. The corresponding rms background
values were 41 \mjya\ and 26 \mjya. The extent of the long axis enclosed by
the 2$\sigma$ contours of the 12.5 \mic\ emission is \app\ 1$''$.64 (100 AU).
At 17.9 \mic , elongation at the 2$\sigma$ level
is \app2$''$.3 (140 AU) with the SE side clearly more extended 
than the NW side. 
 

Available flux densities for 49 Cet at mid-infrared and longer 
wavelengths are listed in Table~1.  
A 200 \mic\ flux density from ISOPHOT
observations (Walker \& Heinrichsen 2000) 
was suspiciously low ($0.32$ Jy) compared to the
value implied by an interpolation between the
100 \mic\ IRAS and 1.2 mm IRAM observations. 
We have therefore listed only the values 
obtained directly from the Infrared Space Observatory (ISO) 
Data Archive  ($1.1\pm0.5$ Jy at 170 \mic\
and $0.75\pm0.5$ Jy at 150 \mic). 
Photometry in U,B,V,J,H,K,L and M bands 
from Sylvester et al.\ (1996) were used to constrain
stellar photospheric properties as outlined below.


\section {Modeling} 

We use a simple model of a flat optically thin 
disk with relatively few parameters (Backman, Gillett
\& Witteborn 1992; Koerner et al.\ 1998; Wahhaj et al.\ 2003; 2005)
to derive the properties of the 49 Cet disk, since our images
have limited spatial resolution, and there are relatively few flux
density measurements. We assume that thermal radiation from 
an annulus of width $dr$ at a distance $r$ from a star is given by:

$$ f_t(r) =  \sigma(r) 
\varepsilon_\lambda \  B[T_p(r),\lambda] \biggl( {2 \pi r dr \over D^2}
\biggr ), $$

\noindent
where  $\sigma(r)$ is the fractional surface density, $T_p(r)$ the grain
temperature, $B[T_p(r),\lambda]$ the Planck function, 
and $D$ the distance. For 49 Cet, $D =$ 61 pc. 
Moderately absorbing dielectrics with 
an effective grain radius, $a$, have radiative efficiency 
$\varepsilon_\lambda = 1.5a/\lambda$ for $\lambda > 1.5a$ and 
$\varepsilon_\lambda = 1$ for $\lambda < 1.5a$ (Greenberg 
1979). The grain temperature for 
efficient absorbers and inefficient emitters  is $T_p(r) = 432 
a_{\mu m}^{-0.2}({L_*/{\rm L_\odot}})^{0.2}(r_{\rm AU})^{-0.4}$ K can be calculated from radiative 
balance equations. For very small grains, $a < 0.05$ \mic ,
which are both inefficient absorbers and emitters,
$T_p(r) = 636 ({L_*/{\rm L_\odot}})^{2/11}(r_{\rm AU})^{-4/11}({T_*}/\Tsun)^{3/11}$ K 
(Backman \& Paresce 1993), where $L_*$ is the  stellar 
luminosity, 18.4 ${\rm L_\odot}$ for 49 Cet. The two 
temperature laws converge at $a = 0.05$ \mic, and varying 
the grain radius across this point does not result in a 
discontinuity in temperature. However, the latter law yields 
degenerate solutions between $\sigma(r)$ and $a$, and so 
only the product of the two quantities can be constrained. 
This degeneracy finds natural expression in the probability 
distributions generated by the Bayesian method which we discuss 
later. The assumed power-law emissivities are admittedly crude 
approximations for the grain temperature behavior. However, 
the constraining power of the extant data is also limited. 
Thus, more realistic temperature laws are counter-productive 
since they unnecessarily complicate the modeling method.
The modeling constraints yielded on grain radii suggests 
a range of temperatures. It is better to work from these 
temperature ranges and investigate what kinds of dust grains 
they are consistent with, under more realistic modeling assumptions. 

To describe the disk morphology and
orientation we introduce parameters \rin , the inner radius,
\delr , the width, $\gamma$ , assuming radial density profile, 
$\sigma$ \app\ $r^{-\gamma}$, $\phi$, the inclination of the disk 
to line of sight (90\dg\ is edge-on) and $\theta$, the PA of long axis. 
We also include an asymmetry parameter $\beta$, 
the factor by which emission from the NW ansa 
exceeds that from the SE. 
We simulate the stellar photosphere by fitting a Kurucz (1993) model
of a star with T$_{eff}$=9250 K to the optical and infrared
photometry of Sylvester et al.\ (1996), and use 
the mid-infrared and longer 
wavelengths estimates from Table~1.  The
resulting photospheric flux density is added
to the central pixel of the model image, and
the final image is convolved with the appropriate PSF star
to construct the simulated emission map. Models are
varied over all parameter space, and the probability
of each model calculated from $\chi^2$ following 
Wahhaj et al.\ (2005).

Initially, the two Keck/MIRLIN images were treated separately in
the model-fitting process to differentiate 
their separate roles in constraining disk parameter values.
Orientation parameters like $\phi$ and $\theta$ 
can be estimated only by fitting to these individual images. 
Most of the emission appears to arise in 
a small inner disk of very small ($a < 1$ \mic )
hot grains that contribute neglibly to the spectral energy
distribution (SED). Both the temperature and the size of
grains are mainly constrained by the relative brightness of the disk 
at $\lambda =$ 12.5 and 17.9 \mic.

\begin{figure}
\centering{ \hbox { 
\includegraphics[height=8cm]{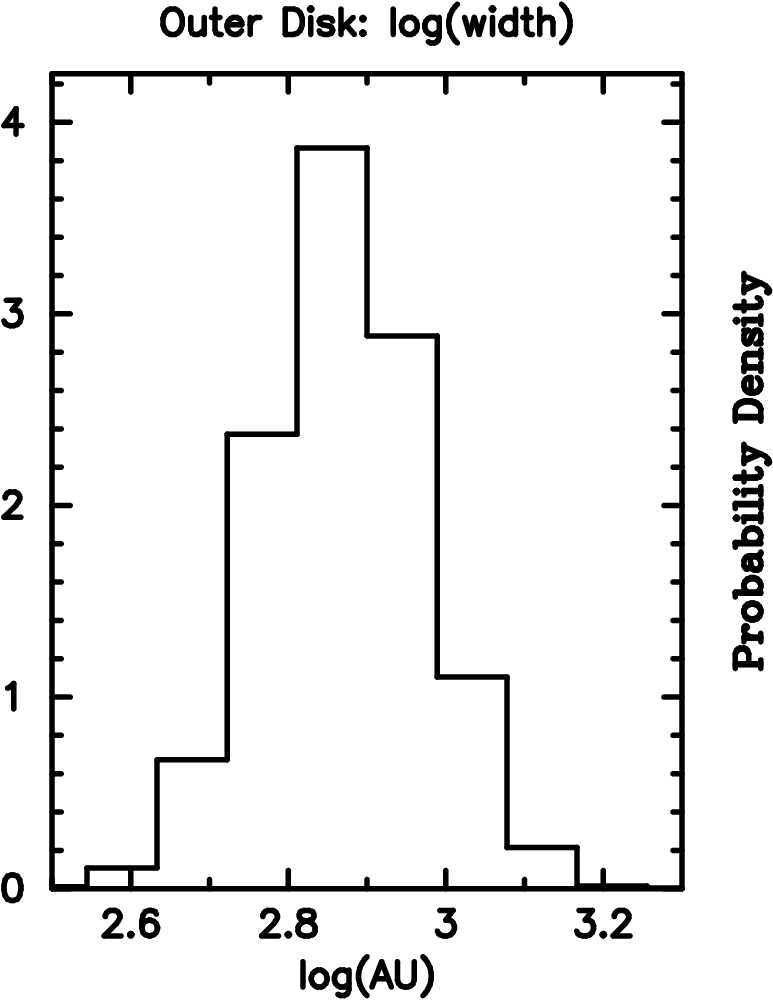}	
\includegraphics[height=8cm]{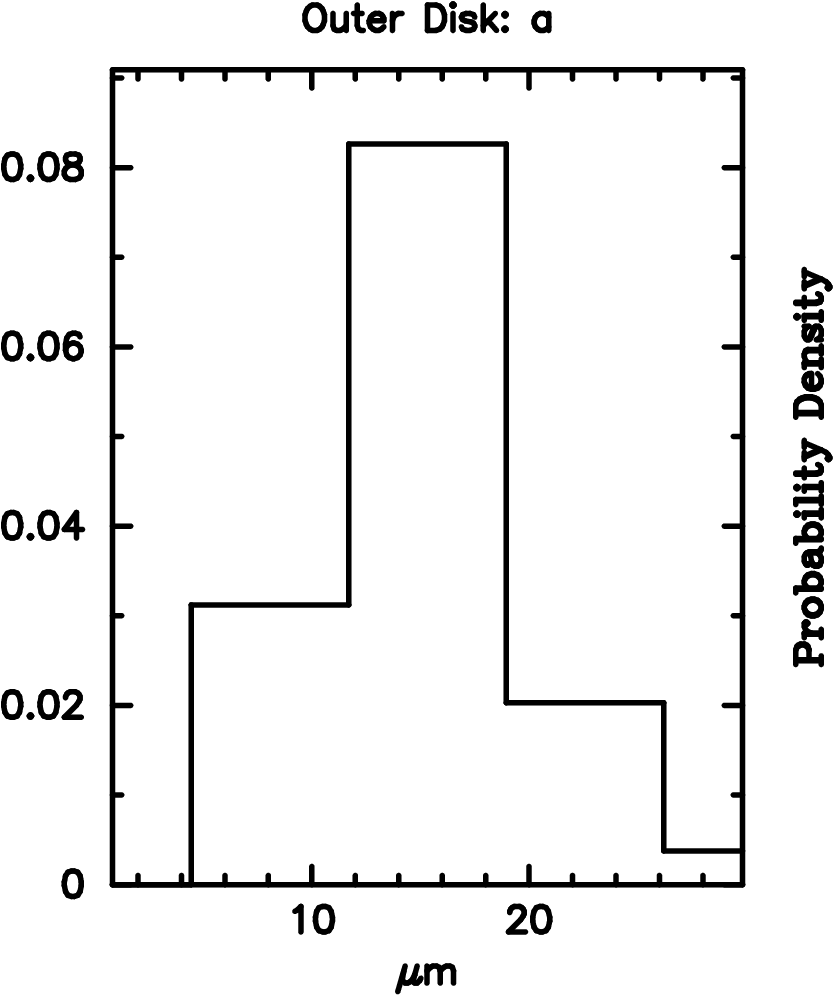}	
} }
\caption{
Probability distributions for selected outer-disk model parameters. 
}
\label{probs2}
\end{figure}

Our subsequent efforts to fit the 
SED and mid-infrared images simultaneously
with a single disk model
failed to reproduce the extensions in 
emission in the mid-infrared images.
Unreduced $\chi^2$ values for the best fit
were \app45, with cardinality of the data only 27 
(see also in Wahhaj et al.\ 2005).  
Fits to the SED alone required a 
much larger disk than that mandated
by the images to reproduce
the long wavelength flux densities and 
yielded only weak constraints
on \rin and grain radius $a$.
We then invoked 
a 2-part disk model comprising 
an inner disk of small grains to simulate the MIR images 
and a large outer disk with a population of large dust
grains to fit the SED. To introduce as few parameters 
as possible the inner radius of the outer disk is 
set to begin at the outer edge of the inner disk. 
These two data sets are complimentary 
in that the 17.9 \mic\ image constrains the inner
radius of the outer disk, while the SED limits 
the inner radius of the inner disk.
This model effectively reproduces all the data,  
and yields a $\chi^2$ of 28.
Most probable parameter values for the inner disk are
$\phi = 55\pm20$\dg , $\theta = 120\pm15$\dg , \rin = $30\pm10$ AU,
\delr = $30\pm20$ AU, $a < 1.0$ \mic , 
$\sigma = 0.1 - 1.5 \times 10^{-3}$,
$\beta = 0.5 - 2$, with no constraints on $\gamma$. 
The probability distribution for the grain radius, $a$ 
actually peaks around $0.1$\mic . The distribution 
suggests no lower limit mostly because the temperature 
law becomes degenerate between grain radius and optical 
depth for $a < 0.5$\mic as discussed earlier. 
Outer disk properties are 
$\phi = 60\pm15$\dg , $\theta = 125\pm10$\dg , 
\rin = $60\pm15$ AU, \delr = $900\pm400$ AU, 
$a = 15\pm10$ \mic , $\sigma = 6.5\pm1.5 \times 10^{-4}$,
$\beta = 1 - 10$, and $\gamma= 0\pm0.5$.
There exists an unrefereed publication of an 850\mic\ JCMT/SCUBA 
detection of 49 Cet (Song, Sandel \& Friberg 2004). At $8.2\pm1.9$mJy, this 
value is inconsistent with the $1.2$mm IRAM detection. We note 
that if the IRAM flux were ignored and 
instead the JCMT flux were used, the outer 
disk width would have to be \app\ 300 AU.
The weak constraints on the disk asymmetry parameters can 
be explained by the fact that the average signal from 
the outer regions of the disk is roughly 3$\sigma$. 
Thus any asymmetry seen in our images are statistically 
not very significant.   
The probability distributions for the inner/outer disk 
configurations are dispayed in Fig.~\ref{probs1}~\&~\ref{probs2}  . 

\begin{figure}
  \centerline{
    \includegraphics[height=8cm]{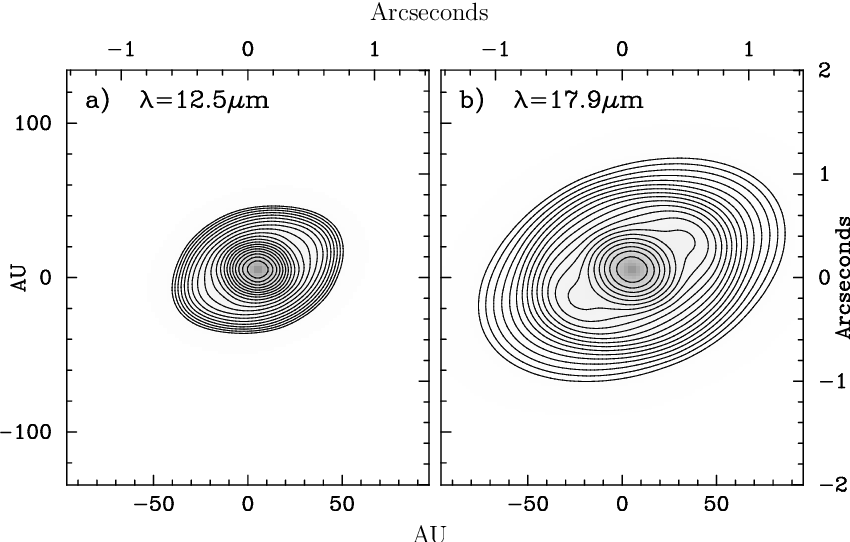}	
  }
  \caption{  Simulated images at 12.5 \mic\ and 17.9 \mic\ from 
	the best-fit model. Contour levels are the same as in Fig.~1. 
        For this model, $\phi = 57$\dg , 
	$\theta = 125$\dg , \rin = $35$ AU,
	\delr = $10$ AU, $a = 0.1$ \mic , and  
	$\sigma = 1.3 \times 10^{-3}$
	for the inner disk, and
	$\phi = 60$\dg , $\theta = 125$\dg , 
        \rin = $45$ AU, \delr = $900$ AU, 
        $a = 23$ \mic , and $\sigma = 5.4 \times 10^{-4}$
	for the outer disk.  
	Since only very weak constraints were found for the $\gamma$ and 
	$\beta$ parameters they were set to the nominal values of 0 and 
      	1 respectively. 
	}
  \label{model}
\end{figure}

\begin{figure}
  \centerline{
    \includegraphics[height=8cm]{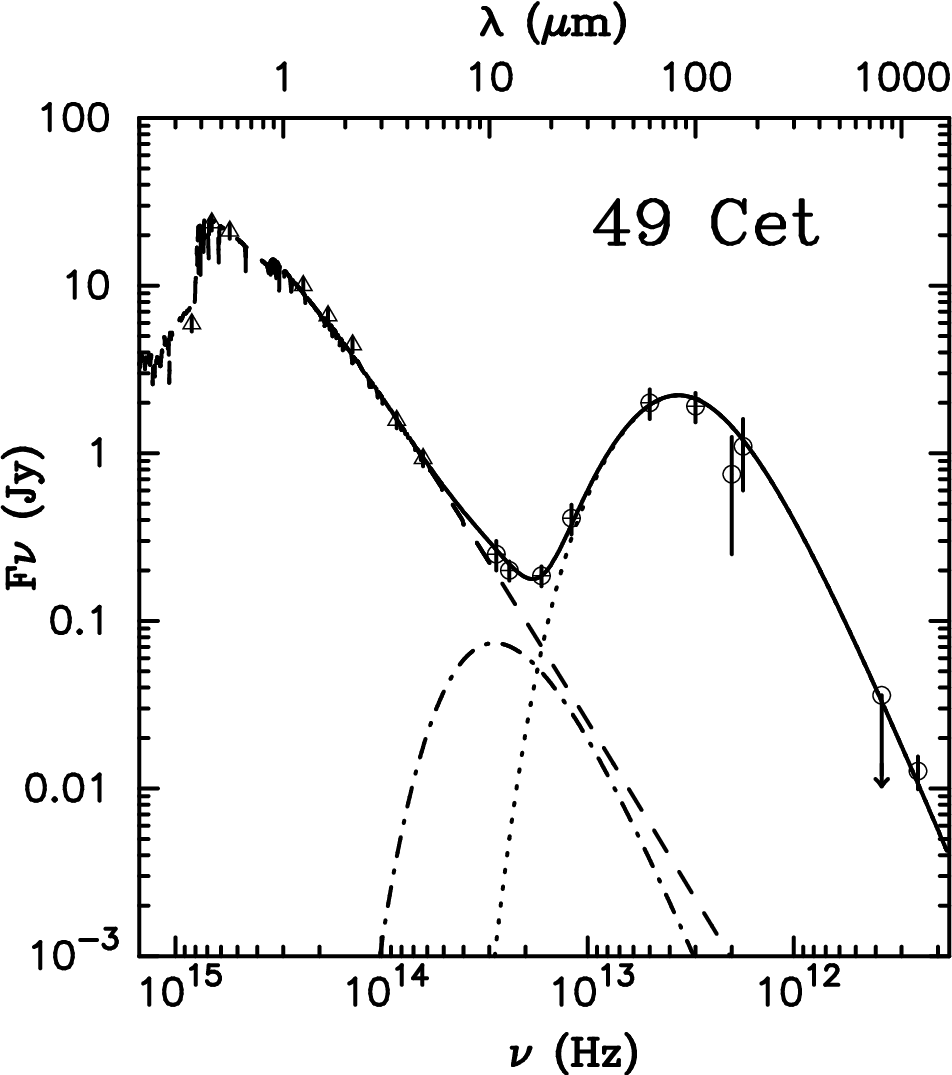}	
  }
  \caption{  Flux densities of 49 Cet from Table.\ 1 (circles) and 
	UBVJHKL \& M photometry from Sylvester et al.\ 1996 (triangles). 
	The solid line represents the complete SED for
	the best-fit model shown in Fig.~\ref{model}. The dashed
	line displays the contribution from the photosphere derived
	from a 1993 Kurucz model fit to the short-wavelength 
	photometry. The dash-dot line is the contribution
	from the inner disk suggested by the Keck MIR images. 
	The dotted line is contribution from
	the outer disk.}
  \label{sed}
\end{figure}

Images simulated from the two-part disk model fit to Keck images
and the SED for 49 Cet are displayed  in Fig.~\ref{model}. 
The simulated SED is shown in Fig.~\ref{sed}.  
We recognize that even this model may be an approximation to some
more complicated structure that is not discernable from the 
current data set. Nevertheless, a number of useful
properties emerge : 1) the 2-component disk 
has a uniform orientation, within uncertainties;  
2) characteristic temperatures derived from a 
grain size-temperature relation (see Backman \& Paresce 1993) 
are \app360 K for inner dust grains
and \app95 K for the outer;
3) the total cross-sectional area of grains in the inner
disk is \app5 AU$^2$, and  
in the  outer disk, \app1650 AU$^2$; 
4) a grain density of $\rho = 2500 {\rm \  kg \ m^{-3}}$ (Jura 1998), 
leads to masses of $6.5 \times 10^{-6}$ \mearth\ and 
$0.32$ \mearth\ respectively. 

\section {Discussion}



\begin{figure}
\centering{ \vbox{
\hbox { 
\includegraphics[height=8cm]{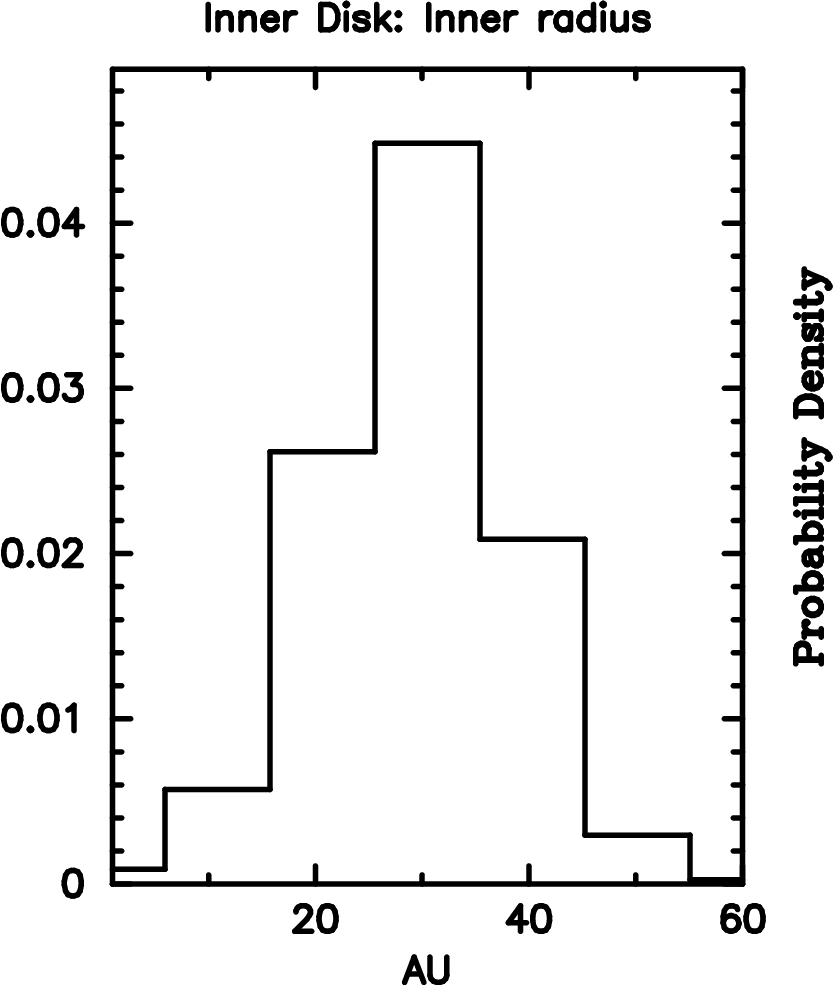}	
\includegraphics[height=8cm]{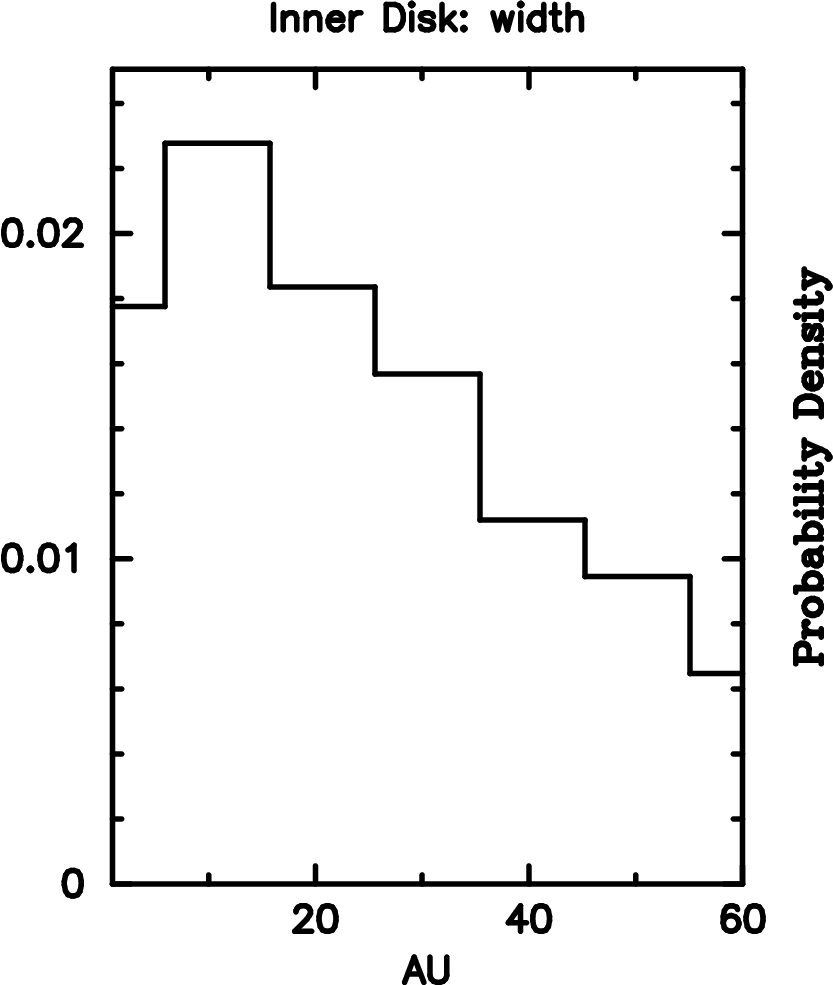}	
}
\hbox {
\includegraphics[height=8cm]{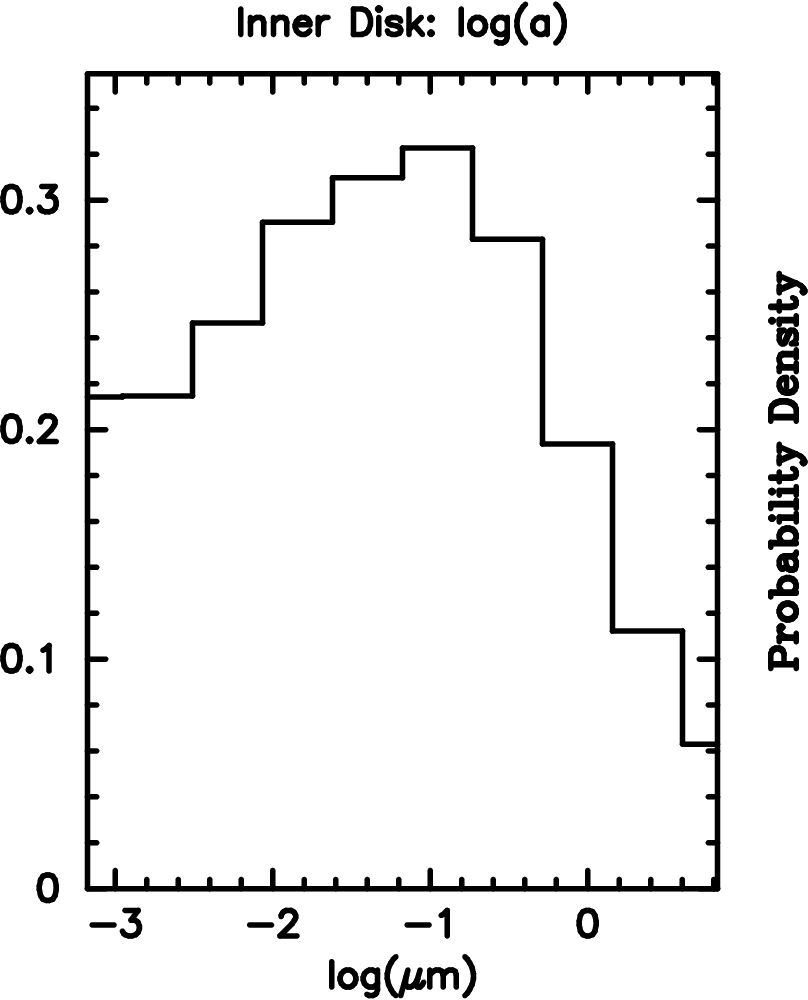}	
\includegraphics[height=8cm]{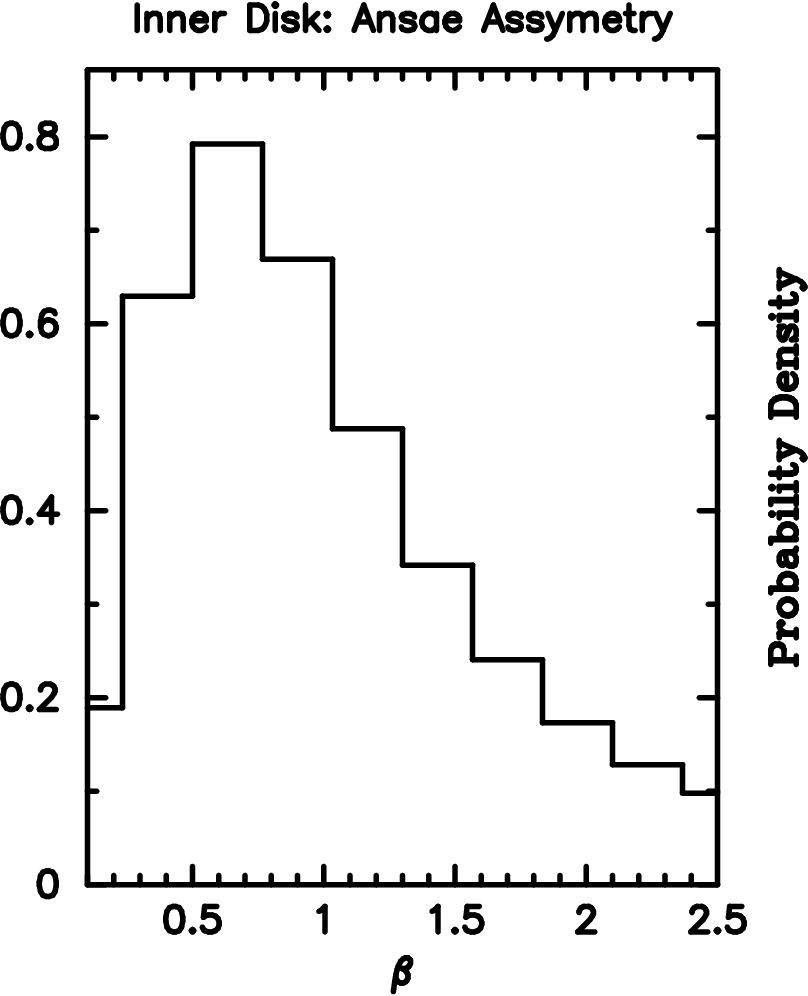}	
} }  
}
\caption{
Probability distributions for selected inner-disk model parameters. 
Most probable values correspond to peaks in the probability distributions. The 
uncertainties are estimated as the 66\% confidence intervals (where our distributions
are binned finely enough) or the shortest range of parameter values that 
encompasses 66\% of the total probability. The most probable values and 
uncertainties for these parameters are detailed in the text.}
\label{probs1}

\end{figure}

Mid-infrared images have also identified 
variations in radial structure in the 
$\beta$ Pic and HR 4796A disks. 
Structures with sharply varying
density and grain size are apparent within the depleted
inner region of $\beta$ Pic's disk 
(Lagage \& Pantin 1994; Backman, Gillett \& Witteborn 1992; 
Wahhaj et al.\ 2003; Weinberger et al.\ 2003; Okamoto et al.\ 2004; 
Telesco et al.\ 2005).
Likewise, a comparison of mid-infrared and Hubble images of 
HR 4796A demonstrates that it is surrounded by a narrow ring 
of large grains embedded
within a wider ring of smaller grains (Wahhaj et al.\ 2005),
properties that are consistent with dispersal of an exo-Kuiper
belt by radiation pressure.
The analysis presented here
suggests a similar morphology for 49 Cet. 
It is evident in  Fig.~\ref{probs1} that there
is only a very small probability that the disk around 49 Cet 
extends all the way to the star. Indeed, it is most likely that the 
region interior to 20 AU is strongly depleted of dust. 
The absense of silicate emission in {\it Spitzer Space Telescope} 
observations of 49 Cet suggests that the circumstellar dust 
must be cooler than 200 K (Kessler-silacci et al.\ 2005). 
In combination with the constraints on grain size obtained here, 
this further strengthens the case for a large inner clearing of 
the disk. 

The major contribution to the excess luminosity comes from
a large-grained outer disk with an inner edge that can barely be 
discerned in the 17.9 \mic\ image. Both the 12.5 and 17.9 \mic\ images 
predominantly trace a region that is approximately 30 to 60 AU
from the star and composed of very small dust grains, 
$a$ \app\ 0.1 \mic . How can dust grains as small as 0.1 \mic\ survive 
in the harsh radiation environment of 49 Cet?
A conceivable explanation can be found in a paper by Artymowicz (1997).
Grain-grain collisions result in a size distribution, $n \sim\ a^{-3.5}$
(Dohnanyi 1969). Grains smaller than \app7 \mic, similar to the 
blow-out ($\beta$-meteoroids) size for  HR 4796A (Wahhaj et al.\ 2005), 
exit the system under radiation pressure. However, as they travel 
outward along the disk mid-plane, they undergo further collisions 
producing a dust avalanche of even smaller grains. 
The resulting amplification in the dust production 
rate depends on the disk surface density and the width of the disk traversed. 
The smallest grains 
(called $\gamma$-meteoroids), will experience 
a relative pile-up, since their outflow is braked more efficiently by residual 
gas (Thebault \& Augereau 2006). An approximate expression for dust production-amplification due 
to the avalanche phenonemon is $e^{N\sigma_m}$, where $N$ is the number 
of particles produced per collision and $\sigma_m$ is the fractional surface 
density experienced by a particle traveling along the disk midplane. 
Collision physics predicts values between 10$^2$ and 10$^3$ for $N$. 
Assuming that the disk thickness is $r/10$, and the inner disk is 
30 AU wide, $\sigma_m\ =\ 8\times 10^{-3}$. Thus the amplification 
factor ranges from 2 to 3000. Detailed numerical simulations of the avalanche 
mechanism show that, in a debris disk with  properties 
similar to 49~Cet, an amplification factor of 200 
may be expected (Grigorieva, Artymowicz \& Thebault\ 2007). Obviously, too high a dust production 
rate predicts too short a disk lifetime. At the same time, rates 
consistent with the pile-up of $\gamma$-meteoroids are accomodated 
by the range of predictions.


 
Another plausible mechanism for $\gamma$-meteoroid production 
is the sublimation of icy bodies
analagous to comets or Kuiper Belt Objects in our own solar
system. Submicron refractory particles may be held together 
by volatile ices to produce large particles unaffected by 
radiation pressure. Upon sublimation, however, the smaller 
particles may be released and subsequently pushed outwards.
Indeed, transient red-shifted lines in the $\beta$ Pictoris
disks have been attributed to falling evaporating bodies (FEBs; 
Beust et al.\ 1990, 1996, 1998). While these events occur only for
bodies that pass within 0.4 AU of the star, 
sublimation of comets and asteroids can
occur as far out as 4 AU (Flammer et al.\ 1998).  
Since the dust/gas ratio in comets is close 
to 1 (Greenberg 1998), large amounts of evaporated 
fine dust particles might exist 
in the region of 49 Cet's inner disk.
Given the wide-ranging grain sizes in optically thin
circumstellar disks, from submicron-sized grains in 49 Cet and 
HD 141569 (Marsh et al.\ 2002) to the 50\mic\ grains in
part of the disk around  HR 4796A (Wahhaj et al.\ 2005),
we conclude that multiple processes may control the evolution of
dust grains both in size and radial location.

In summary our analysis of 49 Cet,
$\beta$ Pic and HR 4796A demonstrates that 
their disks have more in common
than just a large fractional luminosity.  Morphological 
resemblances that set these objects apart from other 
``Vega-like'' A stars, including $\alpha$ Psa and Vega 
itself may be due to their youth, 8-20 Myr 
(see discussion in Jura et al.\ 1993). They may in 
fact be in a transitional stage
of evolution between that of viscous accretion disks around 
Herbig Ae stars (e.g., Mannings \& Sargent 1997; Mannings, 
Koerner \& Sargent 1997) and truly ``Vega-like'' stars with ages
of order 100 Myr or greater. As such, they are interesting 
candidates for constraints on the survival times of circumstellar
gas. Marginal detection of molecular gas (\app\ $10^{-3}$ \msun ) has been reported  
around 49 Cet (Thi et al.\ 2001) based on 
observations of the pure rotational transitions of $H_2$.  
If circumstellar exists, it will be an important factor in the
evolution of dust grains, since this amount of gas 
will shorten the natural lifetime even for grains as large as
a \app\ 50 \mic\ (Klahr \& Lin 2001).  A marginal detection of CO(2$\rightarrow$1) 
emission at 1mm was also reported by Zuckerman et al.\ (1995). However, observations 
of CO(3$\rightarrow$2) at 850 \mic\ by Coulson et al.\ (1998) suggest
a very low upper limit  on the gas mass in the disk
$< 6 \times 10^{-8}$ \msun . CO is easily 
frozen out onto grains in the colder parts of the disk
(Aikawa et al.\ 1997). 

A better understanding of both
the abundance of molecular gas and the morphology of cold dust 
in the outer disk await study at longer wavelengths with
higher resolution and sensitivity, such as will be available
to the next generation of mm-wave arrays.

\acknowledgments
Observations presented here were obtained at the W.M. Keck
Observatory (WMKO), which is operated as a scientific partnership 
among the California Institute of Technology, the University of California 
and the National Aeronautics and Space Administration.  
The Observatory was made possible by the generous financial support of the
W.M. Keck Foundation.  A great
debt is due, also, to Robert Goodrich and the WMKO summit staff for their 
many hours of assistance in adapting MIRLIN to the Keck II visitor
instrument port. The authors wish also 
to recognize and acknowledge the very significant cultural 
role and reverence that the summit of Mauna Kea has always had within the 
indigenous Hawaiian community.  We are most fortunate to have the 
opportunity to conduct observations from this mountain.

\clearpage

\begin{deluxetable}{rrrrrl}
\tablewidth{0pc}
\tablecaption{Flux Densities for 49 Cet}
\tablehead{
\colhead{$\lambda_{eff}$} & 
\colhead{$\delta\lambda$} & 
\colhead{Flux Density} &
\colhead{Uncertainty} &
\colhead{Photosphere} &
\colhead{Reference} \\ 
\colhead{($\mu$m)} &  
\colhead{($\mu$m)} & 
\colhead{(Jy)} &
\colhead{(Jy)} &  
\colhead{(Jy)} &  
\colhead{} }
\startdata
10.8   & 5.3  & 0.25    & 0.05   & 0.191   & Jayawardhana et al.\ 2001    \\ 
12.5   & 1.2  & 0.2     & 0.026  & 0.143   & This work                   \\
17.9   & 2.0  & 0.186   & 0.025  & 0.069   & ''                          \\
12.0   & 6.5  & 0.33    & 0.066  & 0.154   & IRAS FSC                    \\
25.0   & 11.0 & 0.38    & 0.076  & 0.036   & ''                          \\
60.0   & 40.0 & 2.0     & 0.4    & 0.006   & ''                          \\
100.0  & 37.0 & 1.91    & 0.38   & 0.002   & ''                          \\
150.0  & 40.0 & 0.75    & 0.50   & \app\ 0 & ISO                         \\
170.0  & 50.0 & 1.1     & 0.50   & \app\ 0 & ''                          \\
800.0  & -    & $<$0.036\,(3$\sigma$) & -      & \app\ 0 & Zuckerman \& Becklin 1993   \\ 
1200.0 & -    & 0.0127  & 0.0028 & \app\ 0 & Bockelee-Morvan(IRAM) 1994 
\enddata

\end{deluxetable}

\clearpage

\clearpage

\clearpage

\clearpage

\clearpage

\clearpage

\begin{references}	
\reference{adam87} Adams, F.C., Lada, C.J., Shu, F.H.\ 1987, \apj, 312, 788
\reference{aik97} Aikawa, Y., Umebahashi, T., Nakano, T., Miyama, S.M.\ 1997, \apj, 486, L51
\reference{ard04} Ardila, D. R., Golimowski, D. A., Krist, J. E., Clampin, M., Williams, J. P., Blakeslee, J. P., Ford, H. C., Hartig, G. F., Illingworth, G. D.\ 2004, \apj, 617, 147
\reference{arty97} Artymowicz, P.\ 1997, Annual Review of Earth and Planetary Sciences, 25, 175
\reference{bac92} Backman, D.E., Gillett, F.C., \& Witteborn, F.C.\ 1992, \apj, 385, 670
\reference{bac93} Backman, D.E., \& Paresce, F., 1993, in Protostars \& Planets III, (ed.E.H.Levy \& J.I.Lunine), Tucson: University of Arizona Press, p.1253
\reference{beust1990} Beust, H., Vidal-Madjar, A., Ferlet, R., Lagrange-Henri, A.M.\ 1990, Astronomy and Astrophysics, 236, 202 
\reference{beust1996} Beust, H., Lagrange, A.-M., Plazy, F., Mouillet, D.\ 1996, Astronomy and Astrophysics, 310, 181 
\reference{beust1998} Beust, H., Lagrange, A.-M., Crawford, I.A., Goudard, C., Spyromilio, J., Vidal-Madjar, A.\ 1998, Astronomy and Astrophysics, 338, 1015
\reference{bock1994} Bockelee-Morvan, D,, André, P., Colom, P., Colas, F., Crovisier, J., Despois, D., Jorda, L.\ 1994, Circumstellar Dust Disks and Planet Formation, Proceedings of the 10th IAP Astrophysics Meeting, Institut d'Astrophysique, Paris
\reference{ciez06} Cieza, L. A. et al.\ 2006, in preparation.
\reference{clam03} Clampin, M., Krist, J.E., Ardila, D.R., Golimowski, D.A., Hartig, G.F., Ford, H.C., Illingworth, G.D., Bartko, F., Benítez, N., Blakeslee, J.P., Bouwens, R.J., Broadhurst, T.J., Brown, R.A., Burrows, C.J., Cheng, E.S., Cross, N.J.G., Feldman, P.D., Franx, M., Gronwall, C., Infante, L., Kimble, R.A., Lesser, M.P., Martel, A.R., Menanteau, F., Meurer, G.R., Miley, G.K., Postman, M., Rosati, P., Sirianni, M., Sparks, W.B., Tran, H.D., Tsvetanov, Z.I., White, R.L., Zheng, W.\ 2003, \aj, 126, 385
\reference{coul1998} Coulson, I.M., Walther, D.M., Dent, W.R.F.\ 1998, \mnras, 296, 934
\reference{dohn1969} Dohnanyi, J.W.\ 1969, J. Geophys. Res., 74, 2531
\reference{flam1998} Flammer, K.R., Mendis, D.A., Houpis, H.L.F.\ 1998, \apj, 494, 822
\reference{gre79} Greenberg, J.M.\ 1979, Infrared Astronomy, Proceedings of NATO Advanced Study Institute, ed.G.Setti \& G.G.Fazio (Dordrecht:Reidel), 51
\reference{green1998} Greenberg, J.M.\ 1998, Astronomy and Astrophysics, 330, 375
\reference{gri2007} Grigorieva, A., Artymowicz, P., \& Thebault, Ph.\ 2007, Astronomy and Astrophysics, 461, 537
\reference{heap00} Heap, Sara R., Lindler, Don J., Lanz, Thierry M., Cornett, Robert H., Hubeny, Ivan, Maran, S.P. \& Woodgate, Bruce\ 2000, ApJ, 539, 435
\reference{hol98} Holland, W.S., Greaves, J.S., Zuckerman, B., Webb, R.A., McCarthy, C., Couldon, I.M., Walther, D.M., Dent, W.R.F., Gear, W.K., \& Robson, I.\ 1998, \nat, 392, 788 
\reference{jay2001} Jayawardhana, R., Fisher, R.S., Telesco, C.M., Piña, R.K., Barrado y Navascués, D., Hartmann, L.W., Fazio, G.G.\ 2001, \apj, 122, 2047
\reference{jur93} Jura, M., Zuckerman, B., Becklin, E.E., \& Smith, R.C.1993, \apj, 418, L37
\reference{jur98} Jura, M., Malkan, M., White, R., Telesco, C., Pena \& Fisher, R.S.1998, \apj, 505, 897
\reference{k01} Koerner, D.W., Sargent, A.I. \& Ostroff, N.A.\ 2001, \apj, 560, L181
\reference{k98} Koerner, D.W., Ressler, M.E., Werner, M.W., \& Backman, D.E.,1998, \apjl, 503, L83
\reference{kala05} Kalas, P. et al.\ 2005, Nearby Resolved Debris Disks Conference, STSI
\reference{kala06} Kalas, P., Graham, J.R., Clampin, M.C., Fitzgerald, M.P.\ 2006, \apj, 637, 57
\reference{kess05} Kessler-Silacci, J.E., Hillenbrand, L.A., Blake, G.A., Meyer, M.R.\ 2005, \apj, 622, 404
\reference{kla01} Klahr, H.H., Lin, D.N.C.\ 2001, 554, 1095
\reference{kur93} Kurucz, R.\ 1993, ATLAS9 Stellar Atmosphere Programs and 2 km/s grid.Kurucz CD-ROM No.13.Cambridge, Mass.: Smithsonian Astrophysical Observatory.
\reference{liu04} Liu, M.C.\ 2004, Science, 305, 1442
\reference{laga94} Lagage, P.O., Pantin, E.\ 1994, Nature, 369, 628  
\reference{Man1997} Mannings, V.\ \& Sargent, A.I.\ 1997, \apj, 490, 792
\reference{MKS1997} Mannings, V., Koerner, D.W., \& Sargent, A.I.\ 1997, Nature, 388, 555
\reference{Man1998} Mannings, V., Barlow, M.J.\ 1998, \apj, 497, 330
\reference{marsh2002} Marsh, K.A., Silverstone, M.D., Becklin, E.E., Koerner, D.W., Werner, M.W., Weinberger, A.J., Ressler, M.E.\ 2002, \apj, 573, 425
\reference{moor2002} Moór, A., Ábrahám, P., Derekas, A., Kiss, Cs., Kiss, L. L., Apai, D., Grady, C., Henning, Th.\ 2006, \apj, 644, 525
\reference{okam2004} Okamoto, Y.K., Kataza, H., Honda, M., Yamashita, T., Onaka, T., Watanabe, Jun-ichi, Miyata, T., Sako, S., Fujiyoshi, T., Sakon, I.\ 2004, Nature, 431, 660
\reference{ress94} Ressler, M.E., Werner, M.W., Van Cleve, J., Chou, H.A.\ 1994, Experimental Astronomy, 3, 277
\reference{sad86} Sadakane, K., Nishida, M.\ 1986, Astronomical Society of the Pacific, Publications, 98, 685 
\reference{schn99} Schneider, G., Smith, B.A., Becklin, E.E., Koerner, D.W., Meier, R., Hines, D.C., Lowrance, P.J., Terrile, R.J., Thompson, R.I., Rieke, M.\ 1999, ApJ, 513, 127
\reference{sch05} Schneider, G., Silverstone, M.D., Hines, D.C.\ 2005, 629, 117
\reference{sil00} Silverstone, M.D.\ 2000, PhDT, UCLA
\reference{song04} Song, I., Sandell, G., Friberg, P.\ 2004, Debris Disks and the Formation of Planets: A Symposium in Memory of Fred Gillett, ASP Conference Series, 324, 205
\reference{stau1995} Stauffer, J.R., Hartmann, L.W., Barrado y Navascues, D.\ 1995, \apj, 454, 910
\reference{sylv1996} Sylvester, R.J., Skinner, C.J., Barlow, M.J., Mannings, V.\ 1996, \mnras, 297, 915
\reference{T00} Telesco, C.M., Fisher, R.S., Piña, R.K., Knacke, R.F., Dermott, S.F., Wyatt, M.C., Grogan, K., Holmes, E.K., Ghez, A.M., Prato, L., Hartmann, L.W., \& Jayawardhana, R.\ 2000, \apj, 530, 329
\reference{T05} Telesco, C.M., Fisher, R.S., Wyatt, M.C., Dermott, S.F., Kehoe, T.J.J., Novotny, S., Mariñas, N., Radomski, J.T., Packham, C., De Buizer, J., Hayward, T.L.\ 2005, Nature, 433, 133
\reference{ta01} Takeuchi, T. \& Artymowicz, P.\ 2001, \apj, 557, 990
\reference{the06} Thebault, P., Augereau, J.-C.\ 2006, Astronomy and Astrophysics, to be published.
\reference{thi01} Thi, W. F., van Dishoeck, E. F., Blake, G. A., van Zadelhoff, G. J., Horn, J., Becklin, E. E., Mannings, V., Sargent, A. I., van den Ancker, M. E., Natta, A., Kessler, J.\ 2001, \apj, 561, 1074
\reference{wahhaj03} Wahhaj, Z., Koerner, D.W., Ressler, M.E., Werner, M.W., Backman, D.E., Sargent, A.I.\ 2003, \apj, 584, 27
\reference{wahhaj05} Wahhaj, Z., Koerner, D.W., Backman, D.E., Werner, M.W., Serabyn, E., Ressler, M.E., Lis, D.C.\ 2005, \apj, 618, 385
\reference{walk2000} Walker, H.J., Heinrichsen, I.\ 2000, Icarus, 143, 147
\reference{wein03} Weinberger, A.J., Becklin, E.E., Zuckerman, B.\ 2003, \apj, 584, 33
\reference{wil02} Wilner, D.J., Holman, M.J., Kuchner, M.J., Ho, P.T.P.\ 2002, \apj, 569, 115
\reference{zuck93} Zuckerman, B., Becklin, E.E.\ 1993, \apj, 406, 25
\reference{zuck95} Zuckerman, B., Forveille, T., Kastner, J.H.\ 1995, \nat, 373, 6514, 494
\reference{zuck01} Zuckerman, B.\ 2001, Annual Review of Astronomy and Astrophysics, 39, 549
\reference{zuck04} Zuckerman, B. \& Song, I.\ 2004, \apj, 603, 738
\end{references}
\end{document}